\documentclass[]{spie}  

\usepackage{amssymb}
\usepackage{amsmath,amsfonts,xspace,textcomp,graphicx}
\usepackage[colorlinks=true, allcolors=blue]{hyperref}

\newcommand{\mum}{\mbox{{\usefont{U}{eur}{m}{n}{\char22}}m}\xspace}

\newcommand{\sce}{\mbox{SCE\lowercase{x}AO}\xspace}

\newcommand{\FIG}[3]{\includegraphics[width=#1\linewidth,draft=#2]{#3}}

\title{SCExAO, an instrument with a dual purpose: perform cutting-edge science and develop new technologies}

\author[a]{Julien Lozi}
\author[a,b,c,d]{Olivier Guyon}
\author[e]{Nemanja Jovanovic}
\author[a,f]{Sean Goebel}
\author[a,g]{Prashant Pathak}
\author[a]{Nour Skaf}
\author[a,g]{Ananya Sahoo}
\author[h]{Barnaby Norris}
\author[i]{Frantz Martinache}
\author[i]{Mamadou N'Diaye}
\author[j]{Ben Mazin}
\author[j]{Alex B. Walter}
\author[h]{Peter Tuthill}
\author[a]{Tomoyuki Kudo}
\author[k]{Hajime Kawahara}
\author[l]{Takayuki Kotani}
\author[m]{Michael Ireland}
\author[n]{Nick Cvetojevic}
\author[n]{Elsa Huby}
\author[n]{Sylvestre Lacour}
\author[a,d,n]{S\'{e}bastien Vievard}
\author[o]{Tyler D. Groff}
\author[p]{Jeffrey K. Chilcote}
\author[q]{Jeremy Kasdin}
\author[b]{Justin Knight}
\author[r]{Frans Snik}
\author[r]{David Doelman}
\author[a]{Yosuke Minowa}
\author[a]{Christophe Clergeon}
\author[a]{Naruhisa Takato}
\author[l]{Motohide Tamura}
\author[a]{Thayne Currie}
\author[l]{Hideki Takami}
\author[l]{Masa Hayashi}
\affil[a]{National Astronomical Observatory of Japan, Subaru Telescope, 650 North A$\!$`oh\={o}k\={u} Place, Hilo, HI 96720, U.S.A.}
\affil[b]{College of Optical Sciences, University of Arizona, Tucson, AZ 85721, U.S.A.}
\affil[c]{Jet Propulsion Laboratory, 4800 Oak Grove Drive, MS 183-901, Pasadena, CA 91109, U.S.A.}
\affil[d]{Astrobiology Center of NINS, 2-21-1, Osawa, Mitaka, Tokyo, 181-8588, Japan}
\affil[e]{California Institute of Technology, 1200 E California Blvd, Pasadena, CA 91125, U.S.A.}
\affil[f]{Institute for Astronomy, University of Hawai`i, 640 North A$\!$`oh\={o}k\={u} Place, Hilo, HI 96720, U.S.A.}
\affil[g]{Sokendai, Graduate University for Advanced Studies, Kanagawa Prefecture, Miura District, Hayama, Shonankokusaimura, 240-0193, Japan}
\affil[h]{Department of Physics and Astronomy, Macquarie University, NSW 2109, Australia}
\affil[i]{Universit\'{e} C\^{o}te d'Azur, Observatoire de la C\^{o}te d'Azur, CNRS, Laboratoire Lagrange, 96 Boulevard de l'Observatoire, 06300 Nice, France}
\affil[j]{University of California Santa Barbara, Santa Barbara, CA 93106, U.S.A.}
\affil[k]{Univeristy of Tokyo, Tokyo, Bunkyo, Hongo, 113-8654, Japan}
\affil[l]{National Astronomical Observatory of Japan, , 2-21-1, Osawa, Mitaka, Tokyo, 181-8588, Japan}
\affil[m]{The Australian National University, Canberra ACT 0200, Australia}
\affil[n]{Observatoire de Paris, LESIA, 5 Place Jules Janssen, 92190 Meudon, France}
\affil[o]{Goddard Space Flight Center, 8800 Greenbelt Rd, Greenbelt, MD 20771, U.S.A}
\affil[p]{Stanford University, 450 Serra Mall, Stanford, CA 94305, U.S.A.}
\affil[q]{Princeton University, Princeton, NJ 08544, U.S.A.}
\affil[r]{Leiden Observatory, Leiden University, P.O. Box 9513, 2300 RA Leiden, The Netherlands}

\authorinfo{Further author information: (Send correspondence to J.L.)\\J.L.: E-mail: lozi@naoj.org, Telephone: 1 808 934 5949}

\begin{document} 
\maketitle

\begin{abstract}
The Subaru Coronagraphic Extreme Adaptive Optics (SCExAO) instrument is an extremely modular high-contrast instrument installed on the Subaru telescope in Hawaii. SCExAO has a dual purpose. Its position in the northern hemisphere on a 8-meter telescope makes it a prime instrument for the detection and characterization of exoplanets and stellar environments over a large portion of the sky. In addition, SCExAO's unique design makes it the ideal instrument to test innovative technologies and algorithms quickly in a laboratory setup and subsequently deploy them on-sky. SCExAO benefits from a first stage of wavefront correction with the facility adaptive optics AO188, and splits the 600-2400 nm spectrum towards a variety of modules, in visible and near infrared, optimized for a large range of science cases. The integral field spectrograph CHARIS, with its J, H or K-band high-resolution mode or its broadband low-resolution mode, makes SCExAO a prime instrument for exoplanet detection and characterization. Here we report on the recent developments and scientific results of the SCExAO instrument. Recent upgrades were performed on a number of modules, like the visible polarimetric module VAMPIRES, the high-performance infrared coronagraphs, various wavefront control algorithms, as well as the real-time controller of AO188. The newest addition is the 20k-pixel Microwave Kinetic Inductance Detector (MKIDS) Exoplanet Camera (MEC) that will allow for previously unexplored science and technology developments. MEC, coupled with novel photon-counting speckle control, brings SCExAO closer to the final design of future high-contrast instruments optimized for Giant Segmented Mirror Telescopes (GSMTs). 
\end{abstract}


\section{INTRODUCTION}
\label{sec:intro}  

Direct imaging of exoplanets and extrasolar environments is very challenging, due to the high-resolution and high-contrast required. High-contrast imaging instruments like SPHERE\cite{Beuzit2008} and GPI\cite{Macintosh2014} are now detecting new exoplanets, brown dwarf companions and protoplanetary disks. The Subaru Coronagraphic Extreme Adaptive Optics (\sce) instrument\cite{Jovanovic2015}, installed at the 8-m Subaru telescope on top Maunakea in Hawaii, is such a high-contrast imager.

But contrary to the other instruments, \sce has a unique design and operation, allowing for continuous upgrades with the latest hardware, software and algorithms, while still performing cutting-edge science on-sky\cite{Garcia2017,Currie2017}. The modular design of \sce, based on collaborations with teams around the world (U.S., Japan, France, Australia, Netherland), is a testbed for new technologies that will be essential for the future high-contrast imagers on Giant Segmented Mirror Telescopes (GSMTs) like the Thirty Meter Telescope (TMT)\cite{Macintosh2006}. The long therm goal of \sce is actually to evolve into a first generation visitor instrument on the TMT\cite{Jovanovic2016}.

In this paper, we present the recent and near-future upgrades of \sce that will allow for better detection and characterization of circumstellar environments. First, Sec.~\ref{sec:scexao} presents the current configuration of the bench. Then, Sec.~\ref{sec:cameras} describes the new NIR cameras installed recently on the instrument, and how they will improve the wavefront control. Current and future improvement in polarimetry in visible and near infrared (NIR) are presented in Sec.~\ref{sec:ploarimetry}. Finally, Sec.~\ref{sec:corono} presents a few high-contrast devices ---coronagraphs and pupil remappers--- that will allow for better characterization of exoplanets, as well as single mode fiber injection of the exoplanets light for high-resolution spectrography using the Infrared Doppler Spectrograph (IRD).

\section{ARCHITECTURE OF \sce}
\label{sec:scexao}

\sce has a unique design that allow for testing of a variety of new technologies necessary for future high-contrast instruments. \sce is composed of various modules and science cameras, fed with a suite of dichroics and beamsplitters. The instrument has a bottom bench where the infrared light (IR, 0.9--2.5\mum) is analyzed.

\begin{figure}%
\center
\FIG{0.9}{false}{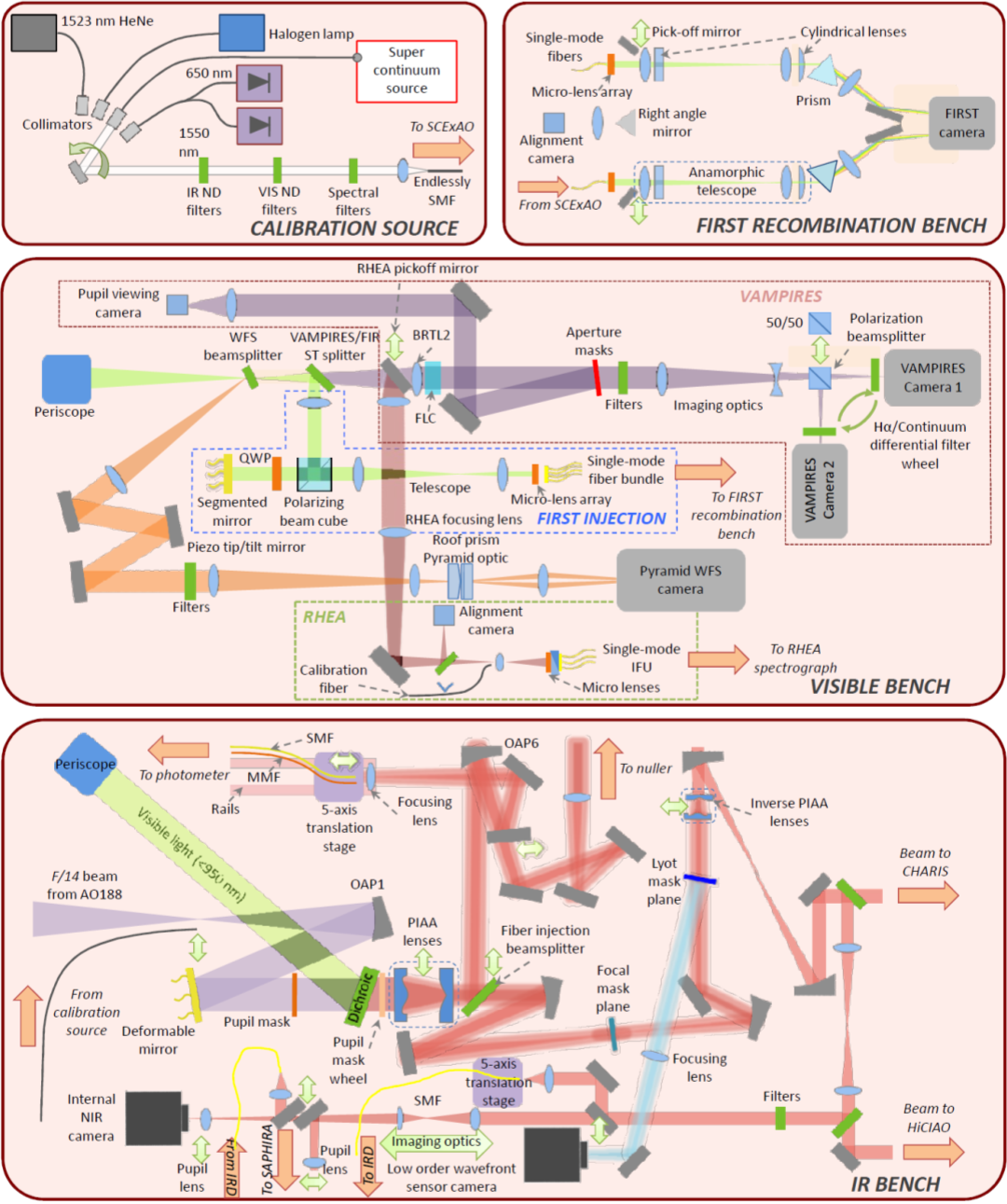}
\caption{Diagram of the \sce instrument.}%
\label{fig:bench}%
\end{figure}

Figure~\ref{fig:bench} presents an updated diagram of the whole instrument from \cite{Jovanovic2015}. The bottom part of the instrument receives the light from AO188, then forms an image on the 2000-actuator deformable mirror (DM). The light is then split between visible and infrared wavelength, at a cutoff of 900~nm. The infrared light goes through a selection of coronagraphs (see Sec.~\ref{sec:corono}). Using beamsplitter wheels, it can then be distributed between various near IR (NIR) cameras. Currently, the main instrument is the integral field spectrograph (IFS) CHARIS\cite{Groff2013} (See Sec.~\ref{sec:charis}. After the decommissioning of HiCIAO, the port is now feeding the new Microwave Kinetic Inductance Detectors (MKIDs) Exoplanet Camera (MEC, See Paper 10702-31 of these proceedings, and Sec.~\ref{sec:mec}). A third port can feed a fast NIR camera, such as the SAPHIRA camera developed at Institute for Astronomy\cite{Atkinson2014} or the First Light C-RED ONE camera, for focal plane wavefront sensing (see Sec.~\ref{sec:lwe}). An internal shortwave IR (SWIR) camera was also recently upgraded to a C-RED2 camera, to perform internal alignments as well as focal plane wavefront sensing for quasi-static low-order aberrations, low-wind/island effect measurement or speckle control (see Sec.~\ref{sec:c-red2}). A post-coronagraphic fiber injection can also be inserted to inject the planet light into a single mode fiber feeding the high-resolution spectrograph IRD\cite{Kotani2014}. Surrounding fibers will select starlight speckles and reimaged on the same internal C-RED2 camera, to perform some wavefront control (see Sec.~\ref{sec:ird}). The NIR light can also be injected in a photonic nulling interferometer (See paper 10701-14).

The visible light reflected off the dichroic is sent to the top bench via a periscope. Once again, beamsplitter wheels split the light between various modules. The higher wavelengths are sent to the visible pyramid Wavefront Sensor (PyWFS). The lower end of the spectrum is split between three modules. VAMPIRES, which uses non-redundant masking with polarimetry to image polarized sources like dust shells and proto-planetary disks\cite{Norris2015}, was upgraded recently (see Sec.~\ref{sec:vampires}). FIRST, a fiber injection interferometric module, can reach milliarcsecond precision in the detection of companions\cite{Huby2012}. Finally, RHEA, a 3x3 fiber fed high-resolution IFS, can resolve convection cells on the surface of the biggest nearby stars\cite{Rains2016}.

\section{NEW CAMERAS FOR IMPROVED WAVEFRONT CONTROL}
\label{sec:cameras}

\subsection{Upgrade to the New C-RED2 Cameras}
\label{sec:c-red2}

The previous internal SWIR camera, with a 170~Hz frame rate and 130~electrons of noise, was upgraded to a First Light Imaging C-RED2 camera. With 400~Hz (600~Hz with the new firmware) and 30~electrons of noise, this camera will allow us to reach fainter stars, and run faster to perform focal plane wavefront control.

\begin{figure}%
\center
\FIG{0.7}{false}{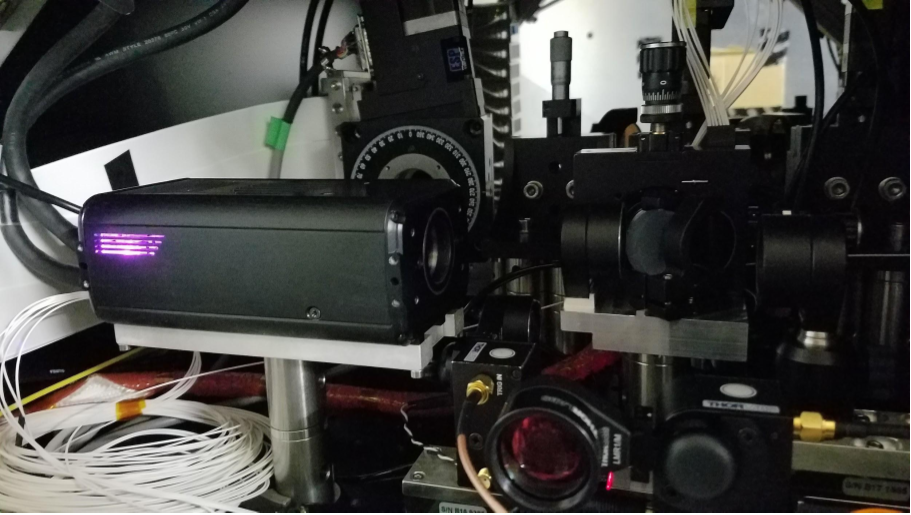}
\caption{New C-RED2 replacing the previous SWIR internal camera inside \sce.}%
\label{fig:cred2}%
\end{figure}

Figure~\ref{fig:cred2} presents a picture of the new C-RED2 camera installed in the bench. The sub-window mode allow for even higher frame rate. The default configuration for this camera is a sub-window using a quarter of the detector, for a field of view of 4x5~arcsec. For this sub-window, the frame rate is 1.2~kHz (1.8~kHz with the new firmware).

This new camera has multiple purposes. It is used to perform alignments, but also various focal plane wavefront controls. A pupil mode is also available to perform precise pupil alignments, and potentially for pupil plane wavefront sensing.

\subsection{Low-Wind Effect/Island Effect Measurement and Control}
\label{sec:lwe}

One of the main unforeseen challenges of high-contrast imaging is the measurement and correction of phase steps behind the telescope spiders. These are likely caused by two different effect (See Paper 10703-83)\cite{Ndiaye2018}:
\begin{itemize}
\item Low-Wind Effect (LWE): the LWE is caused by radiative cooling of the spiders by the sky, which is usually about $40^o$C colder than the ambient temperature at the telescope. The cold spiders then cool down the surrounding air. When wind is low around the spiders, the cold air doesn't get mixed up fast enough with the warmer air, usually on one side of the spiders. A difference of temperature can then be measured between each faces of the spiders, These temperature changes create slow-changing phase gradients and offsets in the pupil. 
\item Island effect: The island effect is a problem created by the PyWFS itself. Phase errors from the turbulence behind the spiders are not measured, and interpolate as continuous phase errors. The PyWFS then can create piston errors between the different segments of the pupil. These errors are much faster than the LWE, usually at the speed of the turbulence. 
\end{itemize}

These two effects produce the same result in the image, i.e. a splitting of the PSF in 2 to 4~lobes of varying brightness.

\begin{figure}%
\center
\FIG{0.5}{false}{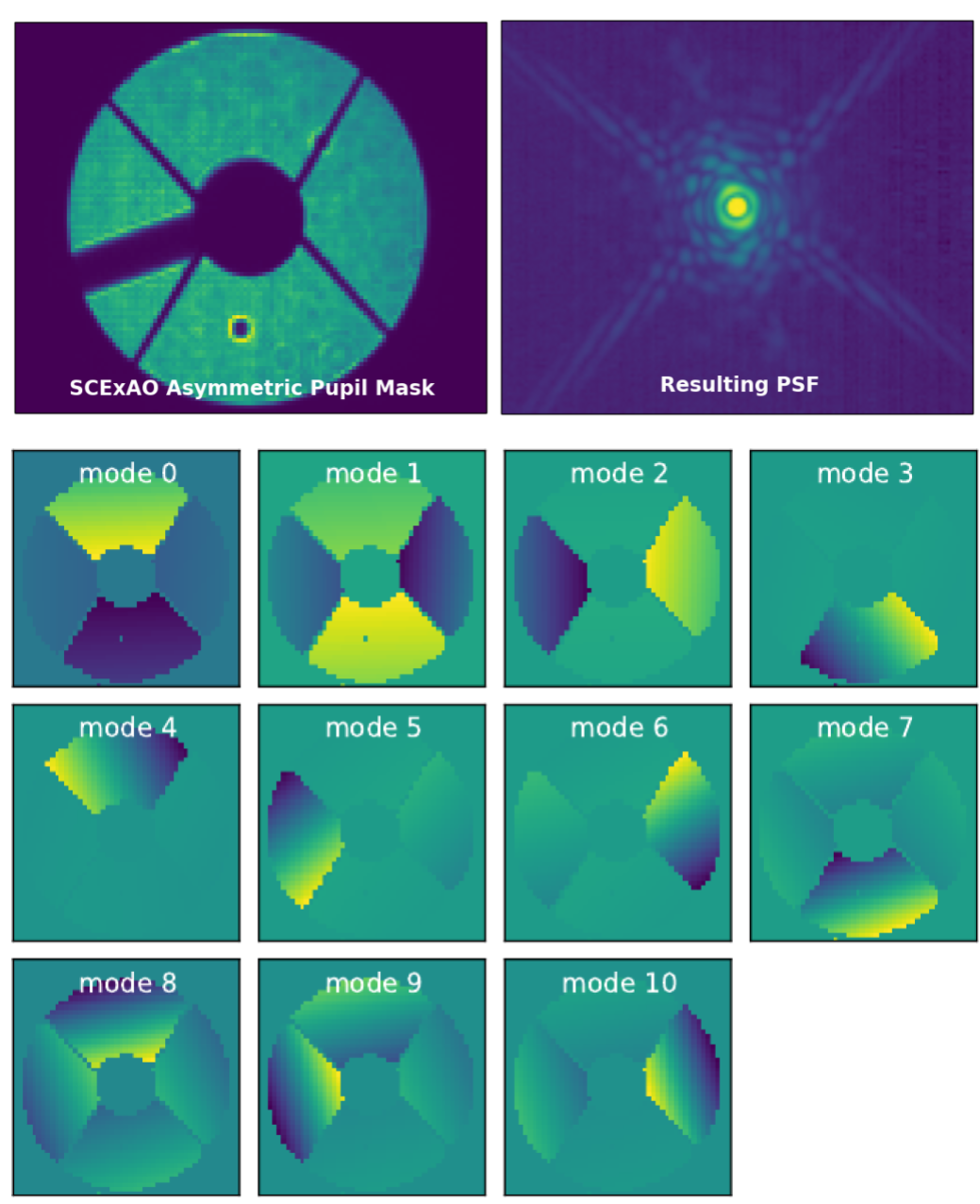}
\caption{Measurement of the LWE modes using the ZAP WFS: pupil image of the asymmetric mask (top left), resulting focal plane image (top right), and orthogonalized basis of LWE modes used for the correction (bottom).}%
\label{fig:lwe}%
\end{figure}

The new C-RED2 will be used to measure these modes using focal plane wavefront imaging. A few options are possible: the Zernike Asymmetric Pupil (ZAP) WFS can measure LWE modes by adding an asymmetry in the pupil\cite{Martinache2013,Martinache2016}. Figure~\ref{fig:lwe} presents the pupil used for ZAP, as well as the focal plane image. Some diffraction is created by the asymmetry. This figure also presents the modes used for the LWE measurements, composed of a orthogonalized base of 11 modes with pistons and tip/tilts on each segment.

The asymmetric mask is unfortunately on the common path to the the NIR science cameras, 
Another focal plane wavefront sensing can be performed using a defocused focal plane image, similarly to the Lyot-stop low-order wavefront sensor (LLOWFS)\cite{Singh2014,Singh2015}.

For more information on low-wind effect and its correction on \sce, see Papers 10703-83 and 10703-49.

\subsection{MKIDS Exoplanet Camera}
\label{sec:mec}

The most important recent upgrade of \sce is the arrival of MEC, a new generation camera that uses the biggest MKIDs detector in the world, with 20,000~pixels. MEC was delivered a few months ago, and is now in a commissioning phase. MEC saw it's first light on-sky at the end of May.

\begin{figure}%
\center
\FIG{0.7}{false}{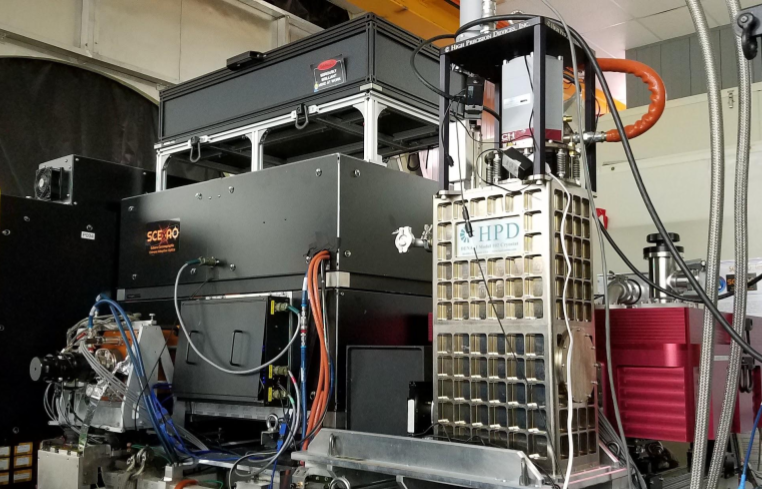}
\caption{Photo of the \sce instrument, with the NIR cameras at the three outputs: from left to right, SAPHIRA, MEC and CHARIS.}%
\label{fig:mkids}%
\end{figure}

Figure~\ref{fig:mkids} presents a picture of the \sce instrument installed  behind AO188, with the three NIR cameras installed on each port: CHARIS (red box on the extreme right), SAPHIRA (orange box on the left) and MEC (silver fridge on the right).

MEC is a noiseless photon-counting camera with a detector cooled at 90~mK. It's properties allow to measure the energy ---therefore the wavelength--- of every photon, with a resolution of about 10. This camera will have a dual purpose: Perform new type of photon-counting science, using for example the statistics of photon arrival time, and perform high-speed focal plane wavefront sensing like speckle control\cite{Martinache2014}. The wavelength information will be useful to correct chromatic effects in the speckle field.

For more information about the MKIDS arrays and MEC, see Paper 10702-31, and speckle nulling with MEC, see Paper 10703-57.

\section{POLARIMETRY}
\label{sec:ploarimetry}

\subsection{New VAMPIRES design with a dual camera system and H-$\alpha$ imaging mode}
\label{sec:vampires}

A major asset in high-contrast imagers is the capacity to analyze polarized sources like protoplanetary disks, or even the reflected light of exoplanets. In visible, the VAMPIRES module performs differential polarization imaging, sometimes coupled with non-redundant masking. When installed in the \sce instrument, VAMPIRES was using a single EMCCD camera with a a slow polarization switching element ($\sim$10~Hz). Last year, VAMPIRES was upgraded to add a second EMCCD and a fast ($\sim$1~kHz) switchable ferroelectric liquid crystal (FLC). Each camera images on linear polarization, while the FLC can switch the polarization state at kHz speed. 

\begin{figure}%
\center
\FIG{0.7}{false}{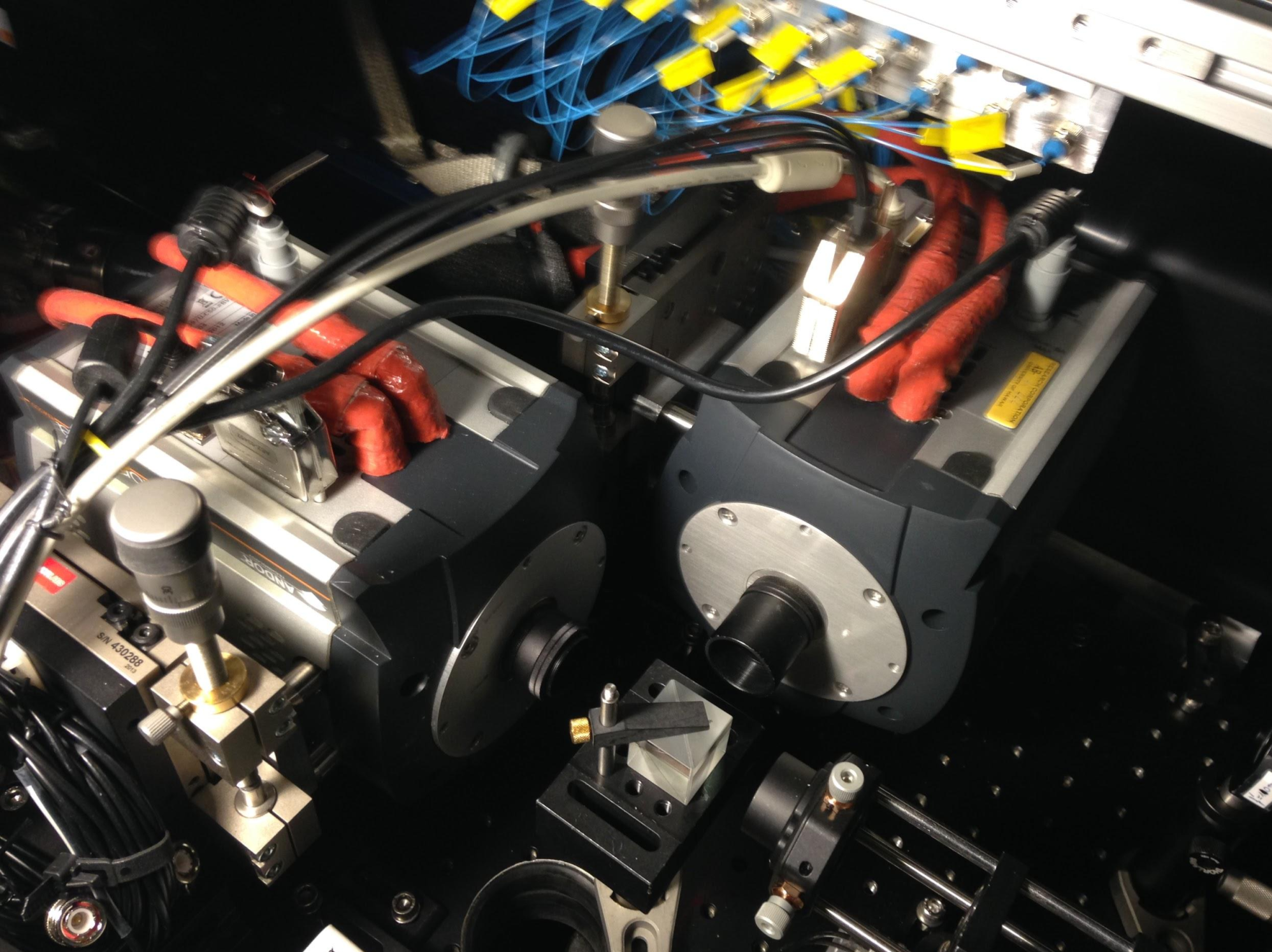}
\caption{New design of VAMPIRES using two EMCCD cameras.}%
\label{fig:vampires}%
\end{figure}

Figure~\ref{fig:vampires} presents the new two-camera system. A polarizing beamcube is splitting the polarization on the two cameras. They are both synchronized to the switching cadence of the FLC. The two-camera system allows now to use small sub-windows to run faster on both cameras.

A new mode was also added: H-$\alpha$ differential imaging. The polarizing beamcube can be replaced by a 50/50 beamcube. A filter wheel common to both cameras can switch H-$\alpha$ and continuum filters in font of the cameras. This new mode will be essential to look at proto-planets that emit a lot of H-$\alpha$ light.

For more information on VAMPIRES, see\cite{Norris2015} and Paper 10702-28.

\subsection{Low-Speed Infrared Polarimetry with CHARIS}
\label{sec:charis}

Other high-contrast instruments like SPHERE and GPI demonstrated the usefulness of infrared polarimetry. We studied possible ways to add a NIR polarization mode inside \sce. The less invasive approach was adopted, with the insertion of a Wollaston prism inside \sce, just before the entrance of CHARIS.

\begin{figure}[b]%
\center
\FIG{0.7}{false}{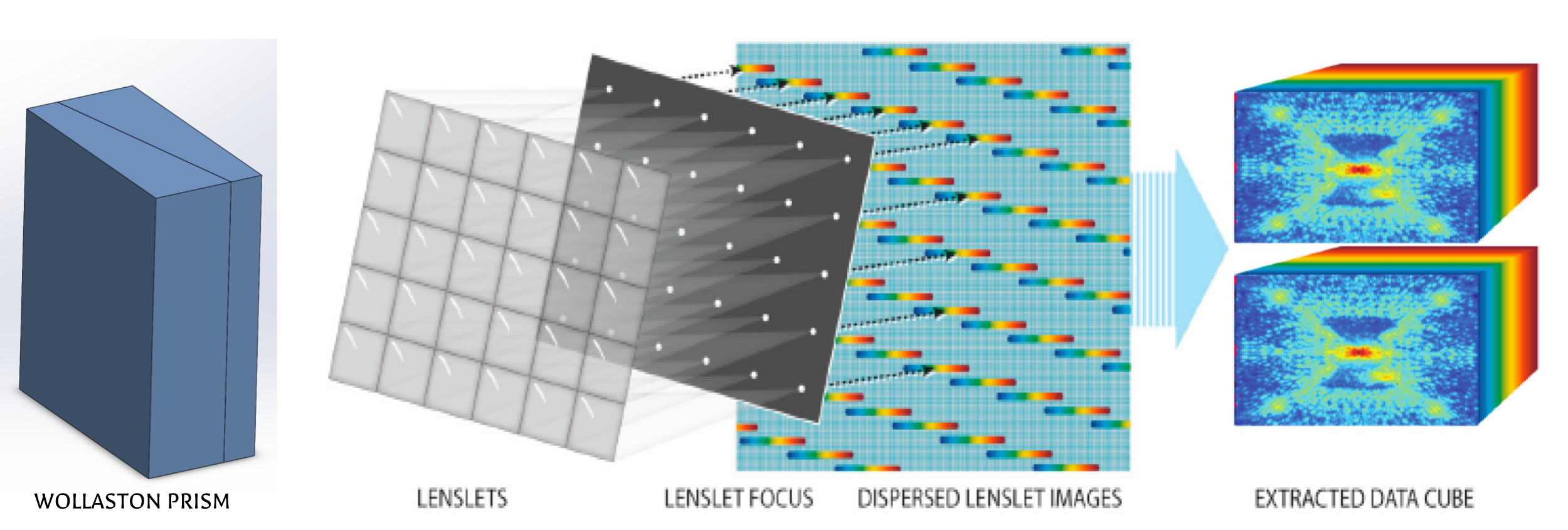}
\caption{Polarizing imaging with the IFS CHARIS, providing at the same time polarization and wavelength information over J- to K-band.}%
\label{fig:charis}%
\end{figure}

Figure~\ref{fig:charis} presents the concept of polarization imaging with CHARIS. The Wollaston prism, placed in a collimated beam just before the pupil plane corresponding to the cold stop of the IFS, sends each polarization at a small angle. The focal plane light is then sampled with a microlens array and an array of field stops. The light of each sample is then dispersed with either a high-resolution prism or a low-resolution prism. The final image is then a data cube with slices corresponding to the different wavelengths between J- and K-band, for both polarizations. Compared to the non-polarized data cube, the only difference is that the field of view of each polarization is divided by half, i.e. $1.1\times2.2$~arcsec. A field stop is also added in the focal plane to avoid cross-talks between the two polarizations. This mode should be completed before the end of 2018.

\subsection{High-Speed Infrared Polarimetry with C-RED2/SAPHIRA}
\label{sec:saphira}

Another polarization mode will be added, in this case without the wavelength information from CHARIS, but with high-speed NIR imaging, with C-RED2 or the SAPHIRA. This mode will be very similar to VAMPIRES, with a fast switching FLC. But contrary to VAMPIRES, only one camera is used, with a Wollaston prism to angularly separate the polarizations on the detector. We will use the same Wollaston design as for CHARIS, and the same field stop in the common path. This mode will also be completed before the end of 2018.

\section{HIGH-CONTRAST IMAGING WITH CORONAGRAPHY AND PUPIL REMAPPING}
\label{sec:corono}

\subsection{Coronagraphs with Small Inner Working Angles}
\label{sec:piaacmc}

One of the main assets of \sce is its capacity to test various coronagraph designs. With a pupil wheel, apodizing lens wheels, a focal plane wheel and a Lyot stop wheel, \sce is now equipped with more than a dozen of coronagraphic options. During science operations, we mostly use a simple Lyot coronagraph with the CHARIS spectrograph, because it is the only broadband option for now. More optimal options in terms of inner working angle (IWA) and contrast are now available but only in H-band. Some will be improved to perform in broadband mode in the near future.

The most anticipated coronagraph installed in \sce is the Phase-Induced Amplitude Apdization Complex Mask Coronagraph (PIAACMC)\cite{Guyon2010a}. A full PIAACMC is now tested for the first time in a high-contrast imager, and shows promising results. It is composed of several elements:
\begin{itemize}
\item Lossless apodizing lenses to apodize slightly the pupil,
\item a complex phase mask in the focal plane, composed of hexagonal segments of various heights to reject the starlight outside the pupil over a large bandwidth,
\item a Lyot stop to block the rejected starlight,
\item inverse apodizing lenses, that remove any off-axis aberrations created by the first lenses.
\end{itemize}

\begin{figure}[b]%
\center
\FIG{0.9}{false}{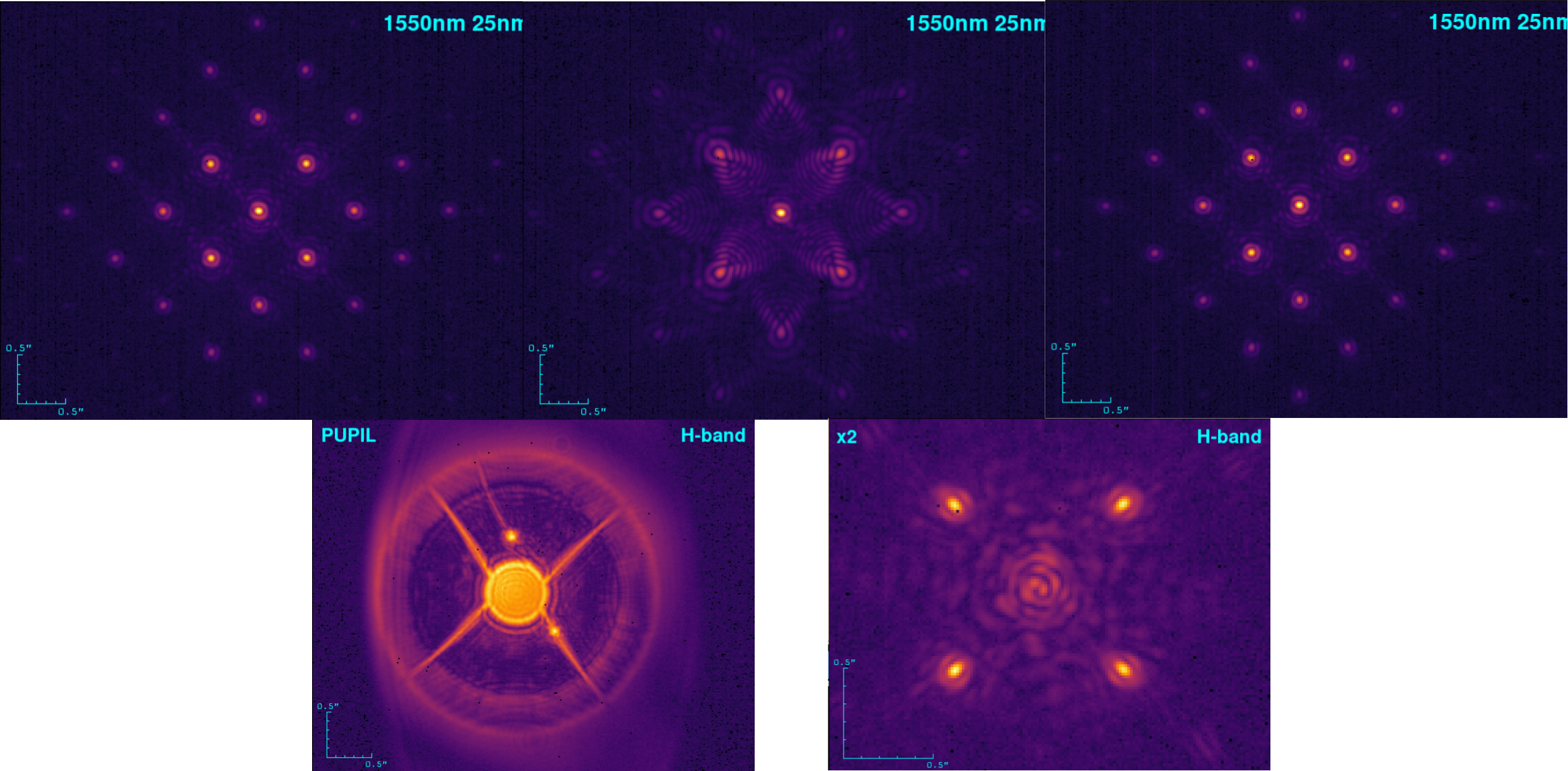}
\caption{PIAACMC lab testing: PSF with speckle grid added using the DM (top left), same speckle grid when the PIAACMC apodizing lenses are inserted (top middle), and when the inverse PIAACMC lenses are inserted (top right). Light rejected in the pupil plane (bottom left) and the focal plane with the full coronagraph, with a fainter speckle grid added for contrast calibration.}%
\label{fig:piaccmc}%
\end{figure}

Figure~\ref{fig:piaccmc} presents lab results with the PIAACMC. When the PIAACMC lenses are inserted, strong aberrations can be seen off-axis, dubbed the pineapple effect. But adding the inverse PIAACMC lenses removes the off-axis aberrations, therefore increase the signal-to-noise ratio (SNR) of stellar companions. 

After the focal plane mask, we can see in Fig.~\ref{fig:piaccmc} that the light is rejected outside the boundaries of the pupil, in the spiders or the central obscuration. The starlight can then easily be suppressed by the Lyot stop, to create the high-contrast image presented in the same figure.

For more details on the characterization of the PIAACMC and the other coronagraphs, see Paper 10706-207. For More details on the design and fabrication of the PIAACMC focal plane mask, see Paper 10706-200.

Lyot stops will soon be replaced with reflective versions, to send the rejected light to the new LLOWS camera, another C-RED2. This new camera will allow for coronagraphic wavefront control on fainter stars and at higher frame rates.

\subsection{Vector Apodizing Phase Plate}
\label{sec:piaacmc}

In addition to coronagraphs, \sce is equipped with pupil plane devices that modify the diffraction pattern to create high-contrast zones around the PSF. A shaped pupil mask creates four high-contrast square zones, and was characterized on-sky.

Another promising high-contrast device is the vector Apodizing Phase Plate (vAPP)\cite{Snik2012}, that uses liquid crystal pattern to create any type of phase aberration like grating, Zernike aberration and diffraction patterns. 

\begin{figure}[b]%
\center
\FIG{0.5}{false}{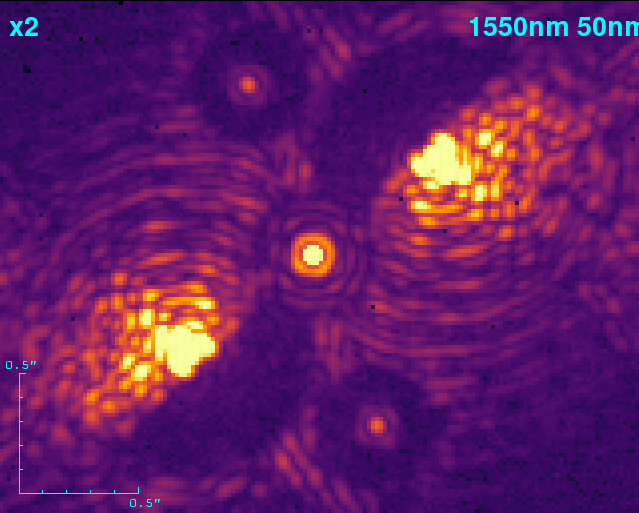}
\caption{Focal plane image after the vAPP, creating symmetric dark zones around copies of the PSF, as well as defocused images for phase diversity measurements.}%
\label{fig:vapp}%
\end{figure}

Figure~\ref{fig:vapp} presents a focal plane image after \sce's vAPP. This device creates two copies of the PSF using a diffraction grating, with symmetric semi-circular dark holes between 2 and 11~$\lambda/D$, as well as two other copies of the PSF with symmetric small focus added, to perform phase diversity. The pattern scales with wavelength, and was designed to fit on the detector of CHARIS between J- and K-band.

For more results on \sce's vAPP, see Paper 10703-8.

\subsection{Post-Coronagraphic Fiber Injection to a Single-Mode Fiber for High-Resolution Spectroscopy}
\label{sec:ird}

High-resolution spectroscopy of exoplanets is a very challenging task. \sce will try to perform this by injecting the post-coronagraphic planet light in a single mode fiber, coupled to the InfraRed Doppler high-resolution spectrograph (IRD)\cite{Kotani2014}.

post-coronagraphic light, combined with speckle suppression loops like speckle nulling, allows to reduce significantly the amount of starlight injected with the planet signal. Another fiber, selecting only starlight, can also be injected in IRD to decipher the planet spectrum from the star spectrum.

\begin{figure}%
\center
\FIG{0.35}{false}{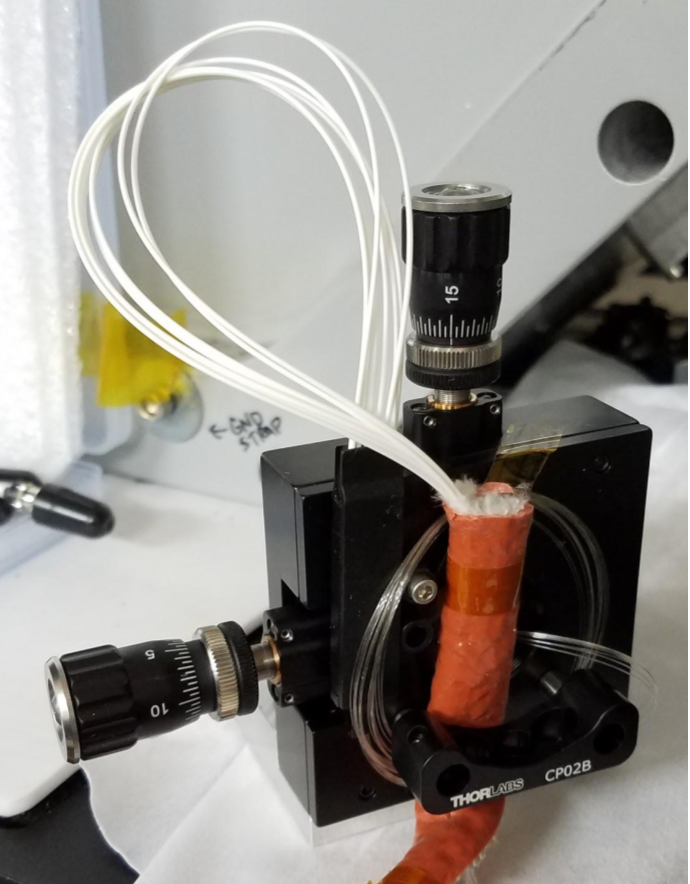}
\caption{V-groove, mounted in a X-Y stage, used to re-image the starlight onto the C-RED2 camera.}%
\label{fig:ird}%
\end{figure}

The light from \sce will actually be injected in a fiber bundle of 7~fibers, a central fiber for the planet light, and 6~surrounding fibers to sample starlight. Since we can only connect two fibers to IRD, the rest of the light is injected into a v-groove, seen in Fig.~\ref{fig:ird}, re-imaged on the internal C-RED2 camera presented in Sec.~\ref{sec:c-red2}. The starlight imaged on the C-RED2 will be used to perform speckle control.

The first test of single mode injection to IRD will happen at the end of June. For more information on IRD, see Paper 10702-37.

\section{CONCLUSION}
\label{sec:conclusion}

\sce is a multi-purpose instrument and testbench, that can perform interesting science at high-contrast and low inner-working angle, and at the same time test cutting edge technologies and algorithms necessary for high-contrast instruments on GSMTs.

\sce is going through continuous upgrades that allow for new and improved science outputs. New high-performance high-speed cameras were recently installed that will provide better focal plane wavefront control. Especially the MKIDS Exoplanet Camera will start a new age in the high-contrast imaging field, with new ways of controlling the speckle field with a small amount of photons.

Polarization is also one of the main focus of the \sce upgrades, with an improved VAMPIRES module in visible. The new two-camera system combined with the FLC allows for faster acquisitions that freeze the speckles. An equivalent system will soon be implemented in NIR, with the new C-RED2 camera and potentially a SAPHIRA detector. A slower polarization will also be implemented in CHARIS, to combine polarization and spectral characterization.

Finally new passive devices like coronagraphs and pupil remappers are now available for science observations, and show already some promising results, in the lab and on-sky. 

\acknowledgments 
 
The development of SCExAO was supported by the Japan Society for the Promotion of Science (Grant-in-Aid for Research \#23340051, \#26220704 \& \#23103002), the Astrobiology Center of the National Institutes of Natural Sciences, Japan, the Mt Cuba Foundation and the directors contingency fund at Subaru Telescope. F. Martinache's work is supported by the ERC award CoG - 683029. The authors wish to recognize and acknowledge the very significant cultural role and reverence that the summit of Maunakea has always had within the indigenous Hawaiian community.  We are most fortunate to have the opportunity to conduct observations from this mountain.

\bibliography{report} 
\bibliographystyle{spiebib} 

\end{document}